\documentclass[%
reprint,
superscriptaddress,
nofootinbib,
 amsmath,amssymb,
 aps,
floatfix,
]{revtex4-2}

\usepackage{graphicx}% Include figure files
\usepackage{dcolumn}% Align table columns on decimal point
\usepackage{bm}% bold math
\usepackage{xcolor}
\begin{document}

\preprint{APS/123-QED}

\title{Electrostatic control of the proximity effect in the bulk of semiconductor-superconductor hybrids}% Force line breaks with \\

\author{N. van Loo$^{\ddagger}$}
\author{G.P. Mazur$^{\ddagger}$}%
\email{mazur.grzesiek@gmail.com}
\author{T. Dvir}
\author{G. Wang}
\author{R.C. Dekker}
\author{J.-Y. Wang}
\author{M. Lemang}
\author{C. Sfiligoj}
\author{A. Bordin}
\author{D. van Driel}
\affiliation{QuTech and Kavli Institute of Nanoscience Delft University of Technology, 2600 GA Delft, The Netherlands}%
\author{G. Badawy}
\author{S. Gazibegovic}
\author{E.P.A.M. Bakkers}
\affiliation{Applied Physics department, 5600 MB Eindhoven, The Netherlands}%
\author{L.P. Kouwenhoven}
\email{l.p.kouwenhoven@tudelft.nl}
\affiliation{QuTech and Kavli Institute of Nanoscience Delft University of Technology, 2600 GA Delft, The Netherlands}%

\date{\today}

\begin{abstract}
The proximity effect in semiconductor-superconductor nanowires is expected to generate an induced gap in the semiconductor. The magnitude of this induced gap, together with the semiconductor properties like the spin-orbit coupling and $g$\,-\,factor, depends on the coupling between the materials. It is predicted that this coupling can be adjusted through the use of electric fields. We study this phenomena in InSb/Al/Pt hybrids using nonlocal spectroscopy. We show that these hybrids can be tuned such that the semiconductor and superconductor are strongly coupled. In this case, the induced gap is similar to the superconducting gap in the Al/Pt shell and closes only at high magnetic fields. In contrast, the coupling can be suppressed which leads to a strong reduction of the induced gap and critical magnetic field. At the crossover between the strong-coupling and weak-coupling regimes, we observe the closing and reopening of the induced gap in the bulk of a nanowire. Contrary to expectations, it is not accompanied by the formation of zero-bias peaks in the local conductance spectra. As a result, this cannot be  attributed conclusively to the anticipated topological phase transition and we discuss possible alternative explanations. 
\end{abstract}

\maketitle
\def\thefootnote{$\ddagger$}\footnotetext{These authors contributed equally to this work}

When a semiconductor is coupled to a superconductor, the resulting hybrid is expected to inherit properties of both. The combination of these properties can be exploited to create exotic phases of matter~\cite{lutchyn:2018_NRP,Prada:2020_NRP}. For example, a magnetic field can trigger the transition to a phase of topological superconductivity in semiconducting nanowires with strong spin-orbit coupling~\cite{Lutchyn_2010:PRL,Oreg_2010:PRL}. In theory, this should be accompanied by the formation of Majorana zero modes (MZMs) at the ends, together with a closing and reopening of the superconducting gap in the bulk of the hybrid~\cite{MSFT_2022:arXiv,pikulin:2021_arxiv}. In general, the proximity effect induces superconductivity in the semiconductor as the result of Andreev reflection at the interface between the materials. This effect manifests itself as the emergence of an induced superconducting gap $\Delta_{\rm {i}}$ in the semiconductor. The size of this gap depends on the size of the proximitizing superconductor $\Delta_{\rm {SC}}$, as well as the coupling between the materials~\cite{shabani_2016:PRB}. Importantly, the coupling also affects various other properties of the hybrid, such as the spin-orbit coupling and $g$\,-\,factor. Moreover, it is expected to be tunable through the use of electric fields~\cite{Antipov:2018_PRX,Reeg_2018:PRB}.

In experiments, tunnelling spectroscopy at the end of a nanowire has been used to demonstrate the tunability of the superconducting gap~\cite{de_Moor_2018:NJP} and the $g$\,-\,factor of Andreev bound states (ABSs)~\cite{Vaitiekenas_2018:PRL,Jiyin_2022:arXiv}. However, such tunnelling experiments only provide information on the local density of states at the end of a nanowire. Yet, it remains unknown what information these observations provide about the proximity effect in the bulk of a hybrid. Advances in nanofabrication now enable the study of semiconductor-superconductor hybrids in a three-terminal geometry~\cite{Menard_2020:PRL,Heedt:2021_NC,Borsoi:2021_AFM}. In addition to the local density of states at the two ends of a nanowire, such devices allow the nonlocal conductance to be measured. Nonlocal transport is possible only in an energy window between the gap of the superconductor and the induced gap in the semiconductor~\cite{Rosdahl:2018_PRB}, and thus can be used to directly determine the induced gap in the bulk of the hybrid~\cite{Anselmetti_2019:PRB}. Measurements in this geometry have been used to observe the closing of the induced gap~\cite{Puglia:2021_PRB}, map the local charge of ABSs~\cite{Danon_2020:PRL,Poschl_2022:arXiv} and investigate the quasiparticle wavefunction composition~\cite{Wang_2021:arXiv}.

In this letter, we investigate the effect of gate-induced electric fields on the bulk of InSb nanowires, proximitized by Al/Pt films~\cite{Mazur_2022:AM}. To do this, we utilize nonlocal spectroscopy. We demonstrate that the devices can be tuned into a strongly-coupled regime with an induced gap close to that of the Al/Pt shell. Likewise, gate voltages can be used to significantly reduce the induced gap and eventually fully close it. By applying a parallel magnetic field, we show that wires in the strong-coupling regime can have critical magnetic fields close to that of the superconducting shell. On the other hand, a gate-reduced coupling drastically lowers the critical field.

\begin{figure}[t]
\includegraphics[width=\columnwidth]{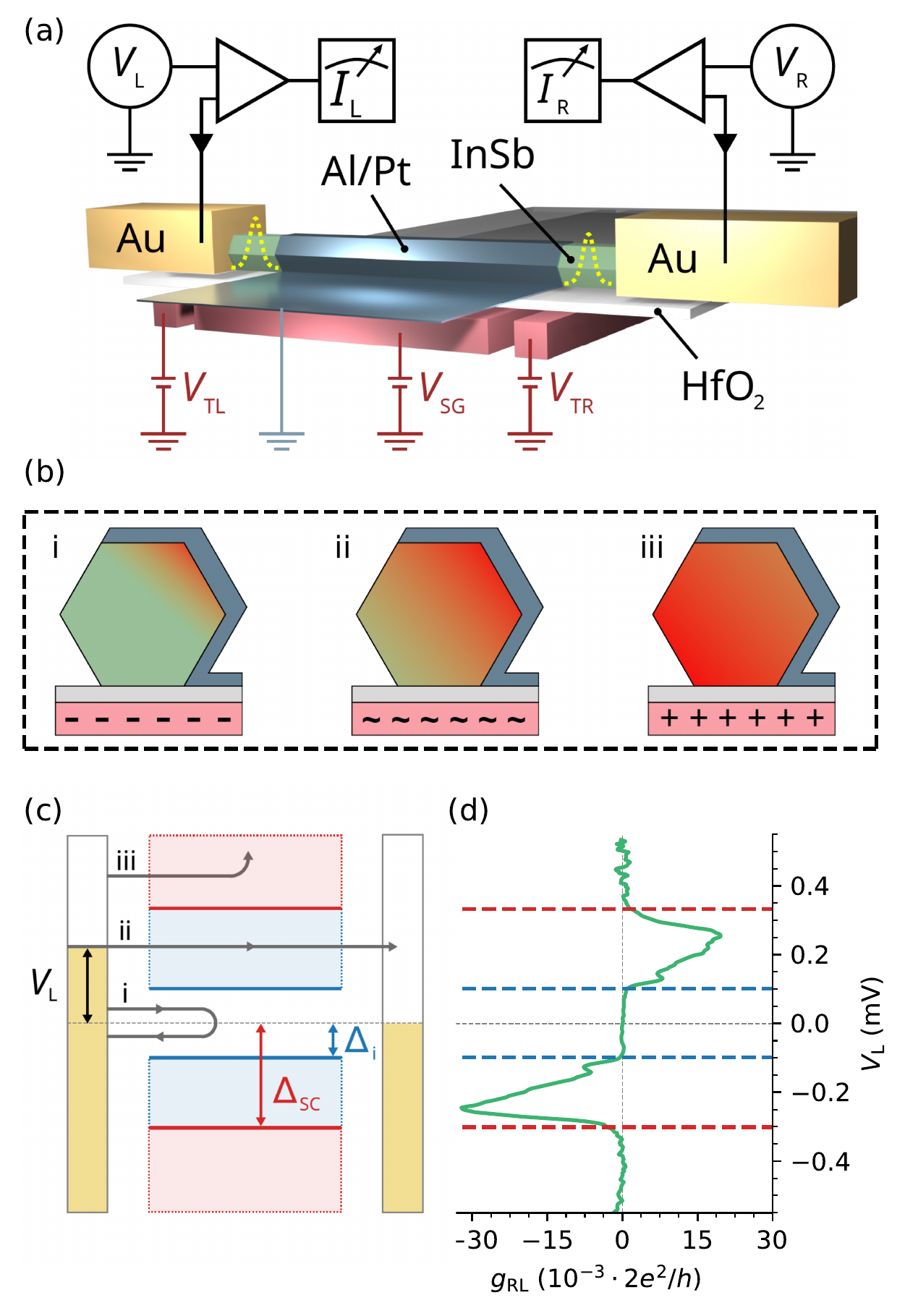}
\caption{\label{fig:1}(a) Schematic of a three-terminal hybrid device and the measurement circuit. The superconducting shell is grounded through its connection to the film on the substrate. Yellow dashed potentials indicate the formation of tunnel barriers in the semiconducting junctions. (b) Illustration of three different coupling regimes between the superconductor and semiconductor. (i) Strong-coupling: electrons (red) are confined at the interface which results in a renormalization of the semiconducting properties. (ii) Intermediate-coupling: predicted to be optimal for the formation of a topological superconductor. (iii): Weak-coupling: electrons accumulate far from the interface, which can result in unproximitized states. (c) Transport schematic of nonlocal measurements. (i) Below $\Delta_{\rm {i}}$ (blue), only local processes are possible. (ii) In between $\Delta_{\rm {i}}$ and $\Delta_{\rm {SC}}$, nonlocal transport can occur. (iii) Above $\Delta_{\rm {SC}}$, electrons are drained to ground. (d) Example of measured nonlocal conductance $g_{\rm {RL}}$ corresponding to the diagram in (c). Blue and red dashed lines indicate the induced and superconductor gap, respectively.}
\end{figure}

The three-terminal devices presented in this work are fabricated using our shadow-wall lithography technique~\cite{Heedt:2021_NC,Borsoi:2021_AFM}. In Fig.~\ref{fig:1}a we depict the device schematic of a nanowire hybrid used in these experiments. A set of pre-patterned bottom gates is separated from the InSb nanowire by a thin layer of HfO$_\mathrm{2}$. Voltages on the two tunnel gates, $V_{\mathrm{TL}}$ and $V_{\mathrm{TR}}$, are used to induce tunnel barriers in the exposed semiconducting segments. The super gate voltage $V_{\mathrm{SG}}$ is used to apply an electric field in the bulk of the hybrid. The nanowire is covered on three facets by an Al/Pt film, where the Pt serves to enhance the critical magnetic field of the Al film~\cite{Mazur_2022:AM}. This superconducting shell extends onto the substrate, forming the connection to ground. Two Cr/Au contacts are fabricated at the ends of the wire. The devices are measured by individually applying bias voltages, $V_{\mathrm{L}}$ and $V_{\mathrm{R}}$, to the left and right leads. The conductance matrix is obtained by measuring the differential conductances $g_{\rm ij}$ $\equiv$ d$I_{\rm i}/$d$V_{\rm j}$, with $i,j = L,R$ using standard lock-in techniques (see Supplementary section I and II for details of device fabrication and measurement details). 

\begin{figure}[b]
\includegraphics[width=\columnwidth]{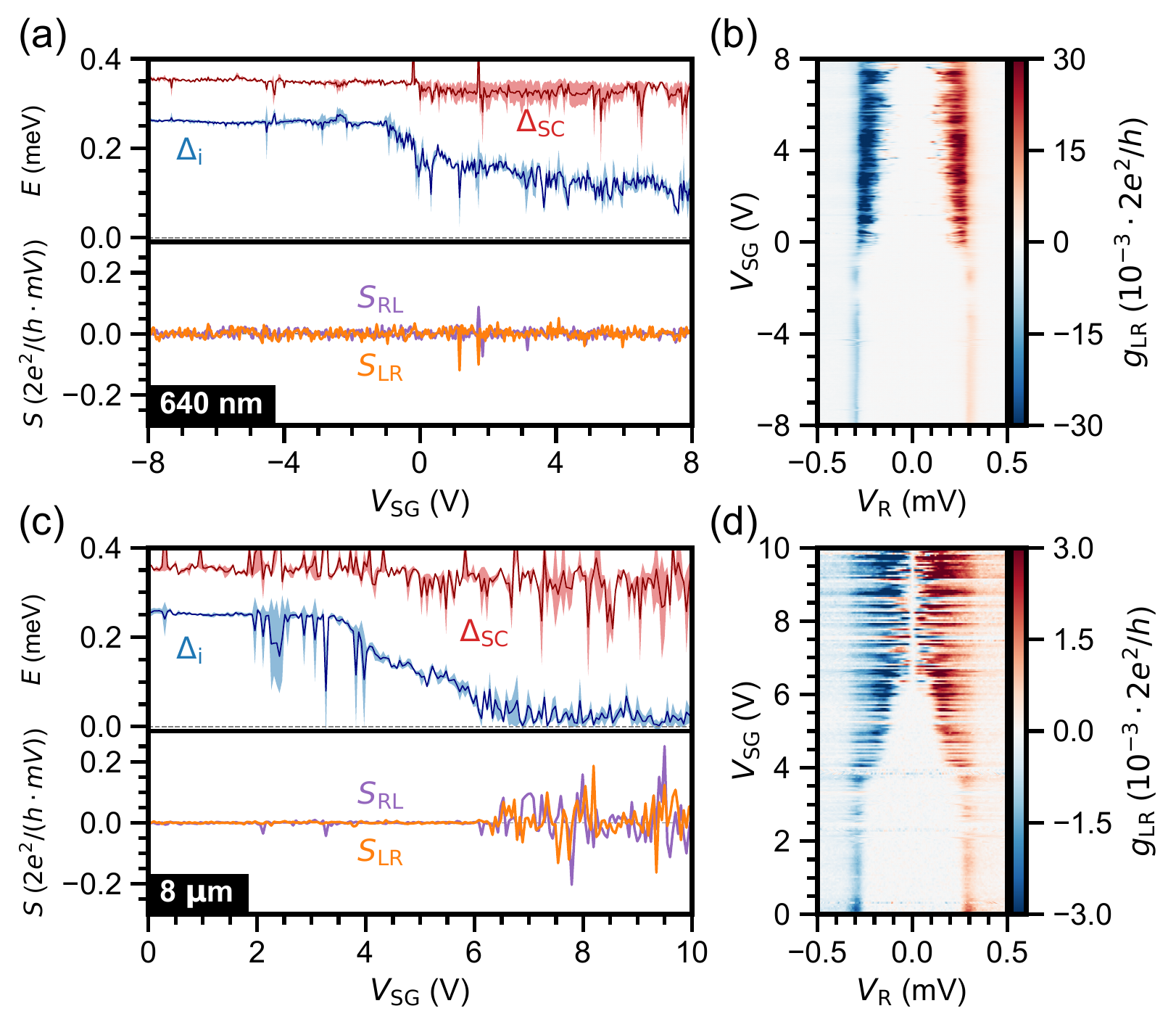}
\caption{\label{fig:2}Nonlocal conductance as a function of $V_{\rm SG}$ in the absence of a magnetic field. (a) Top: $\Delta_{\rm {i}}$ (blue) and $\Delta_{\rm {SC}}$ (red). Dark colors represent the mean of four values, obtained from the positive and negative biases of the two nonlocal signals. Similarly, the shaded areas correspond to the standard deviation. Data taken on device A (640\,nm long hybrid). Bottom: calculated nonlocal slope at zero bias for $g_{\mathrm{RL}}$  (purple) and $g_{\mathrm{LR}}$ (orange). (b) One of the conductance matrix elements $g_{\mathrm{LR}}$ corresponding to (a). (c) and (d) present the gaps, nonlocal slope and the corresponding conductance matrix element similar to (a) and (b) for device B (8\,$\mu$m long hybrid).}
\end{figure}

In Fig.~\ref{fig:1}b, we illustrate the expected effect of electric fields on the bulk of the hybrid as calculated by~\cite{Shen_2021:PRB}. For negative gate voltages (Fig.~\ref{fig:1}b(i)), electrons accumulate near the semiconductor-superconductor interface which results in a strong coupling to the superconductor. As a consequence, the semiconducting properties of the hybrid are strongly renormalized. We refer to this as the strong-coupling regime in the rest of this work. On the other hand, electrons can accumulate far from the interface through the application of positive gate voltages (Fig.~\ref{fig:1}b(iii)). This results in a diminished coupling with unproximitized states in the hybrid, to which we refer as the weak-coupling regime. Finally, there is a crossover between these two regimes (Fig.~\ref{fig:1}b(ii)) where electrons still maintain superconducting correlations, while their semiconducting properties are only moderately renormalized. As a result, this crossover is expected to be optimal for the emergence of topological superconductivity~\cite{Shen_2021:PRB}. Furthermore, the application of an electric field also changes the electron density in the hybrid. Due to quantum confinement we expect the formation of discrete subbands, each with their own coupling strength. Thus, applied gate voltages should be able to tune the hybrid between the different subbands.

\begin{figure}[t]
\includegraphics[width=\columnwidth]{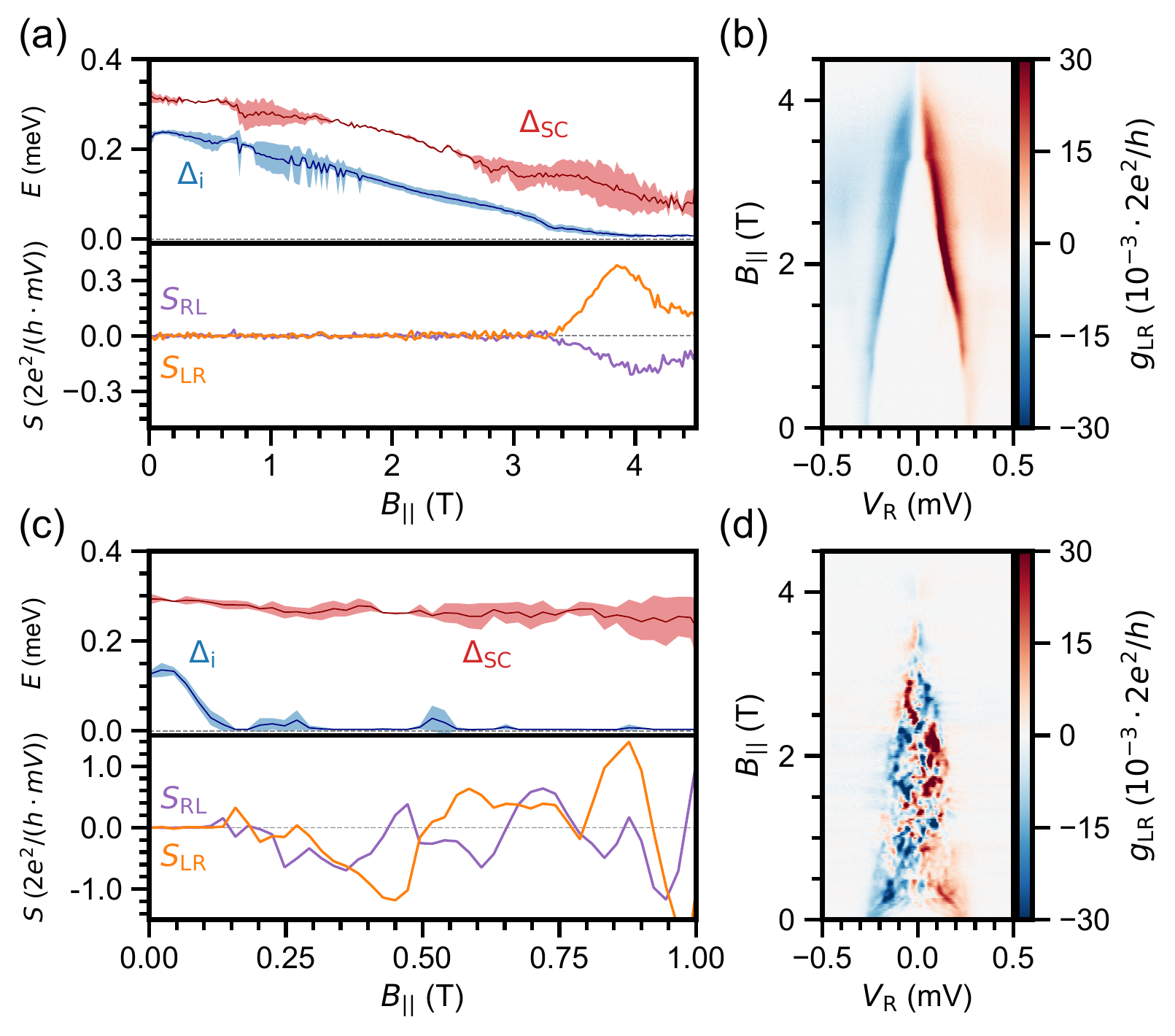}
\caption{\label{fig:3}Nonlocal conductance as a function of $B_{\rm {\parallel}}$ for device C (1 $\mu$m hybrid). (a) Top panel: $\Delta_{\rm {i}}$ (blue) and $\Delta_{\rm {SC}}$ (red) In the strong-coupling regime ($V_{\rm SG} = -0.75$\,V). Bottom panel: Nonlocal slope extracted from $g_{\rm RL}$ (purple) and $g_{\rm LR}$ (orange). (b) One of the conductance matrix elements $g_{\rm {LR}}$ corresponding to (a). (c) and (d) present the gaps, nonlocal slope and the corresponding conductance matrix element similar to (a) and (b) for the weak-coupling regime ($V_{\rm SG} = 0.5$\,V.)}
\end{figure}

To characterize the different coupling regimes, we determine the induced gap in our devices using nonlocal spectroscopy. The transport mechanisms involved in such measurements are schematically depicted in Fig.~\ref{fig:1}c, together with an example of the resulting nonlocal conductance $g_{\rm {RL}}$ in Fig.~\ref{fig:1}d. If the applied bias $V_{\rm {L}}$ is below the induced gap $\Delta_{\rm {i}}$, electrons from the lead can only enter the superconducting region through Andreev reflection (Fig.~\ref{fig:1}c(i)). This results in the formation of Cooper pairs, which drain away into the superconducting lead. As a consequence, no nonlocal conductance is observed below the induced gap (Fig.~\ref{fig:1}d). Similarly, electrons injected above the gap of the superconductor $\Delta_{\rm {SC}}$ are likely to drain to the ground without reaching the other side~\cite{Denisov_2021:SST} (Fig.~\ref{fig:1}c(iii)). However, if the applied bias is larger than $\Delta_{\rm {i}}$ but below $\Delta_{\rm {SC}}$, injected electrons can reach the opposite lead of the device. This results in a finite nonlocal conductance as shown in Fig.~\ref{fig:1}d, from which $\Delta_{\rm {i}}$ (dashed blue lines) and $\Delta_{\rm {SC}}$ (dashed red lines) can be estimated. In Supplementary section III, we describe how these parameters are determined from the data. While this picture helps to understand three-terminal measurements, we note that nonlocal processes can involve energy relaxation of the injected electrons as well as non-equilibrium effects not captured by the single particle transport theory~\cite{Wang_2021:arXiv}. We further elaborate on this in Supplementary section II.B.

First, we investigate the gate tunability of the induced gap. We primarily focus on the nonlocal signals to identify both the induced gap in the bulk, as well as the superconducting gap in the Al/Pt shell. The full conductance matrices corresponding to the measurements are presented in Supplementary section IV. In Fig.~\ref{fig:2}b, we show how the nonlocal signal $g_{\mathrm{LR}}$ responds to the application of a voltage on the super gate for a short nanowire hybrid (device A, 640\,nm). We see that for a large range of negative voltages, the nonlocal spectrum exhibits two sharp anti-symmetric peaks. This indicates that here the difference between $\Delta_{\rm {i}}$ and $\Delta_{\rm {SC}}$ is small, and so it is associated with the strong-coupling regime. Above a certain gate voltage $V_{\rm {SG}} > -1\,V$, the peaks gradually become wider. This signals the reduction of the induced gap, as the coupling between the semiconductor and superconductor is decreased. In Fig.~\ref{fig:2}a, the top panel shows the behavior of the induced gap in the bulk (blue) and the gap of the superconducting shell (red). By tuning the super gate voltage, the induced gap can be reduced to roughly half of its initial value which implies that all semiconductor states that couple to both leads reside above an energy threshold even at strong electric fields. We suspect this is the result of the finite size of the nanowire, where the potential barriers at the ends provide a source of mixing between proximitized and unproximitized states in the hybrid segment~\cite{Shen_2021:PRB}. In addition, we show in the bottom panel the nonlocal slope at zero bias~\cite{Puglia:2021_PRB}. This parameter is defined as $S_{\rm {ij}} \equiv $ d$^{2}I_{\rm i}/$d$V^{2}_{\rm j}|_{V_{j}=0}$, with $S_{\rm {RL}}$ presented in purple and $S_{\rm {LR}}$ in orange. It stays close to zero, indicating that the bulk maintains a finite induced gap. In Fig.~\ref{fig:2}c and d, we show similar data for a long nanowire (device B, 8\,$\mu$m). The induced gap exhibits a similar abrupt reduction above a certain gate voltage, which is associated with a transition between the strong-coupling and weak-coupling regimes. Furthermore, the induced gap can be fully closed for long nanowires. This is signaled by the deviation of the nonlocal slope from zero above $V_{\rm {SG}} > 6\,V$ in the bottom panel of Fig.~\ref{fig:2}c. We generically observe the tunability of the induced gap, and hence the coupling between the semiconductor and superconductor. However, the application of an electric field does not exclusively tune the coupling but also controls the density in the hybrid. Typically, we observe a  sudden onset of the reduction of $\Delta_{\rm {i}}$. Concurrently, the magnitude of the nonlocal signal increases. We suspect that this behavior is related to the occupation of an additional subband with a reduced coupling. Still, it remains unknown how many sub-bands are active in our hybrids.

\begin{figure*}[t]
\includegraphics[width=\textwidth]{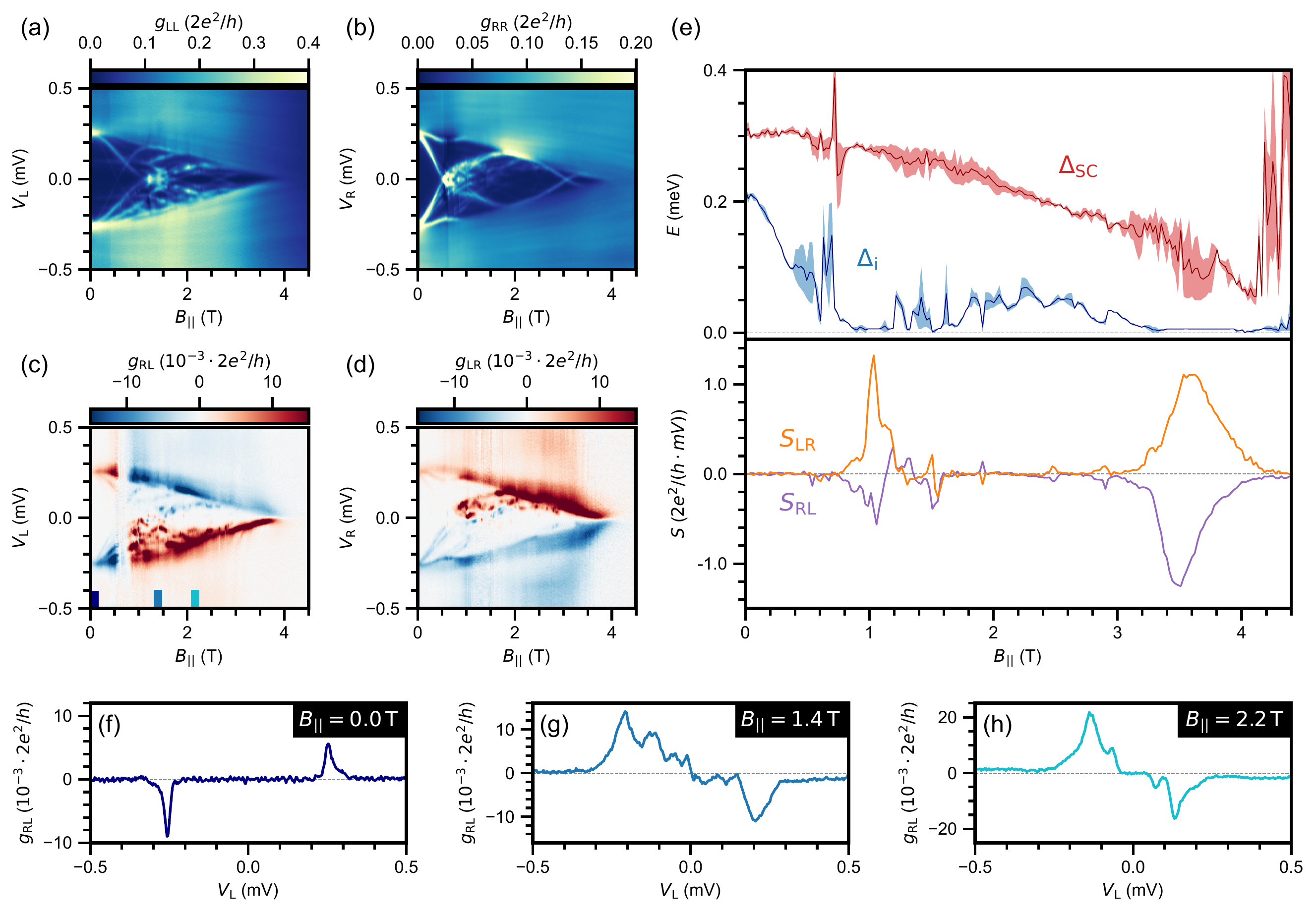}
\caption{\label{fig:5}Nonlocal conductance as a function of $B_{\rm {\parallel}}$ for device C (1 $\mu$m long hybrid). (a-d) Conductance matrix measured as a function of B$_{\parallel}$ at $V_{\rm SG} = -0.3$\,V. (e) $\Delta_{\rm {i}}$ (blue) and $\Delta_{\rm {SC}}$ (red) corresponding to the conductance matrix in (a-d). Bottom: Nonlocal slope extracted from $g_{\rm RL}$ (purple) and $g_{\rm LR}$ (orange). (f-h) Nonlocal conductance $g_{\rm RL}$ presented for (f) B$_{\parallel} = 0$\,T with a large induced gap, (g) $B_{\parallel}=1.4$\,T illustrating a closed induced gap, (h) $B_{\parallel}=2.2$\,T showing flat nonlocal conductance around zero-bias corresponding to a reopening of the induced gap.}
\end{figure*}

We proceed by exploring the effect of parallel magnetic fields $B_{\rm {\parallel}}$ on the induced gap of a 1\,$\mu$m long hybrid (device C). In Fig.~\ref{fig:3}, we present two field sweeps of the nanowire in the two extreme regimes. In the strong-coupling regime (Fig.~\ref{fig:3}a and b), the induced gap (blue) decreases slowly with magnetic field. From both the induced gap and the nonlocal slope, we observe an induced critical field $B_{\rm {\parallel}}^{\rm {c}} = 3.5$\,T. The outer ridge of the nonlocal signal decreases more slowly, which indicates that the shell maintains a superconducting gap (red) up to higher fields. By fitting the linear part of the induced-gap closing to the Zeeman energy $E_{\rm {Z}} = g\mu_{\rm{B}}B/2$, we estimate the $g$\,-\,factor to be $g = 2.3$ (see Supplementary section IV). This demonstrates that the semiconductor properties are indeed strongly renormalized in this regime~\cite{Antipov:2018_PRX,Reeg_2018:PRB}. Such a low $g$\,-\,factor and the absence of any states below $\Delta_{\rm {i}}$ may suggest that the semiconductor is depleted. Yet, we observe that the induced critical field in the strong-coupling regime varies strongly from wire to wire, and likely depends on the microscopic details (see Supplementary section IV).  Moreover, we note that the addition of Pt in the shell causes its $g$\,-\,factor to be reduced close to zero, so that the effective $g$\,-\,factor in the hybrid can be reduced below $g = 2$~\cite{Mazur_2022:AM}. In the weak-coupling regime on the contrary, the induced gap closes at low magnetic fields (Fig.~\ref{fig:3}c and d). From both the induced gap and the nonlocal slope, we observe an induced critical field $B_{\rm {\parallel}}^{\rm {c}} = 0.16$\,T. We estimate a $g$\,-\,factor of $g = 54$, although this value can be overestimated as orbital effects of the magnetic field are more prominent in this regime~\cite{Nijholt_2016:PRB,Winkler_2019:PRB}. The rapid closing of the induced gap confirms that the hybrid inherits more of the semiconductor properties in the weak-coupling regime.

We next turn our attention to the crossover between these two regimes, which is expected to be optimal for the formation of a topological superconducting phase~\cite{Shen_2021:PRB}. In Fig.~\ref{fig:5}a-d, the conductance matrix of the same nanowire (device C, 1\,$\mu$m) taken at  $V_{\rm {SG}} = -0.3\,V$ is shown. In the nonlocal spectra (Fig.~\ref{fig:5}c and d), we see a collection of states moving down in energy as the magnetic field is increased. The induced gap closes around $B_{\rm {\parallel}}^{\rm {c}} = 0.8$\,T and reopens around $B_{\rm {\parallel}}^{\rm {c}} = 1.6$\,T. The induced gap (blue) and nonlocal slope are presented in Fig.~\ref{fig:5}e. Here, the closing and reopening of the induced gap is directly visible. The reopened gap reaches energies of $\Delta_{\rm i} = 50\,\mu$eV, which is similar to predictions of the gap size in topological systems~\cite{pikulin:2021_arxiv}. The reopening is also reflected in the behavior of the nonlocal slope, which deviates from zero around $B_{\rm {\parallel}} = 1$\,T before returning to zero again at higher fields. Fig.~\ref{fig:5}f-h provide linecuts from $g_{\rm {RL}}$, emphasizing that the induced gap is finite at zero field, closed at intermediate field, and reopened at higher fields. However, neither of the local signals (Fig.~\ref{fig:5}a and b) exhibit zero-bias peaks. This suggests that the observed feature does not originate from a topological phase with Majorana zero modes at the ends, extended over the full length of the hybrid. Yet, it may be possible that the presence of tunnel gates generates a smooth potential profile near the ends of the wire. In this case, the local spectra only represent the presence of bound states formed on the smooth potential, while pushing the Majorana zero modes towards the center of the hybrid - effectively decoupling them from the leads~\cite{Woods_2021:PRA,Prada_2012:PRB}. Accordingly, the gap reopening in the bulk should remain visible in the nonlocal spectra as this effectively measures the largest gap in the system, while no zero-bias peaks are observed in the local signals (see Supplementary section IIB). This scenario is supported by the observation that the local signals $g_{\rm {LL}}$ and $g_{\rm {RR}}$ do not appear to depend on the length of the hybrid and are not always correlated, as we elaborate on Supplementary section IV. On the contrary, it is also possible that the reopening of the gap has a topologically trivial origin. For example, the hybrid segment is only 1\,$\mu$m long such that it can be within the short wire limit. This results in a spectrum comprising of discrete energy levels with a small energy spacing, making the concept of topology ill-defined~\cite{Pan:2021_PRB}. Alternatively, the observed gap reopening can originate from two sets of trivial ABSs localized near the nanowire junctions. In this case, spatial overlap due to a long localization length can enable transport through the hybrid~\cite{Hess_2022:PRB}. A similar observation was recently made in phase-biased Josephson junctions~\cite{Banerjee_2022:arXiv_1}. Likewise, orbital effects of the magnetic field may also be responsible for the closing and reopening the induced gap~\cite{Nijholt_2016:PRB}. The most prominent example is the Little-Parks effect, which causes the induced gap to close and reopen once a flux quantum is threaded through the nanowire cross-section~\cite{Valentini_2021:S,Sabonis_2020:PRL}. It has been shown theoretically that such effects can happen in the geometry used in this work~\cite{Winkler_2019:PRB}. 

\begin{figure}[t]
\includegraphics[width=\columnwidth]{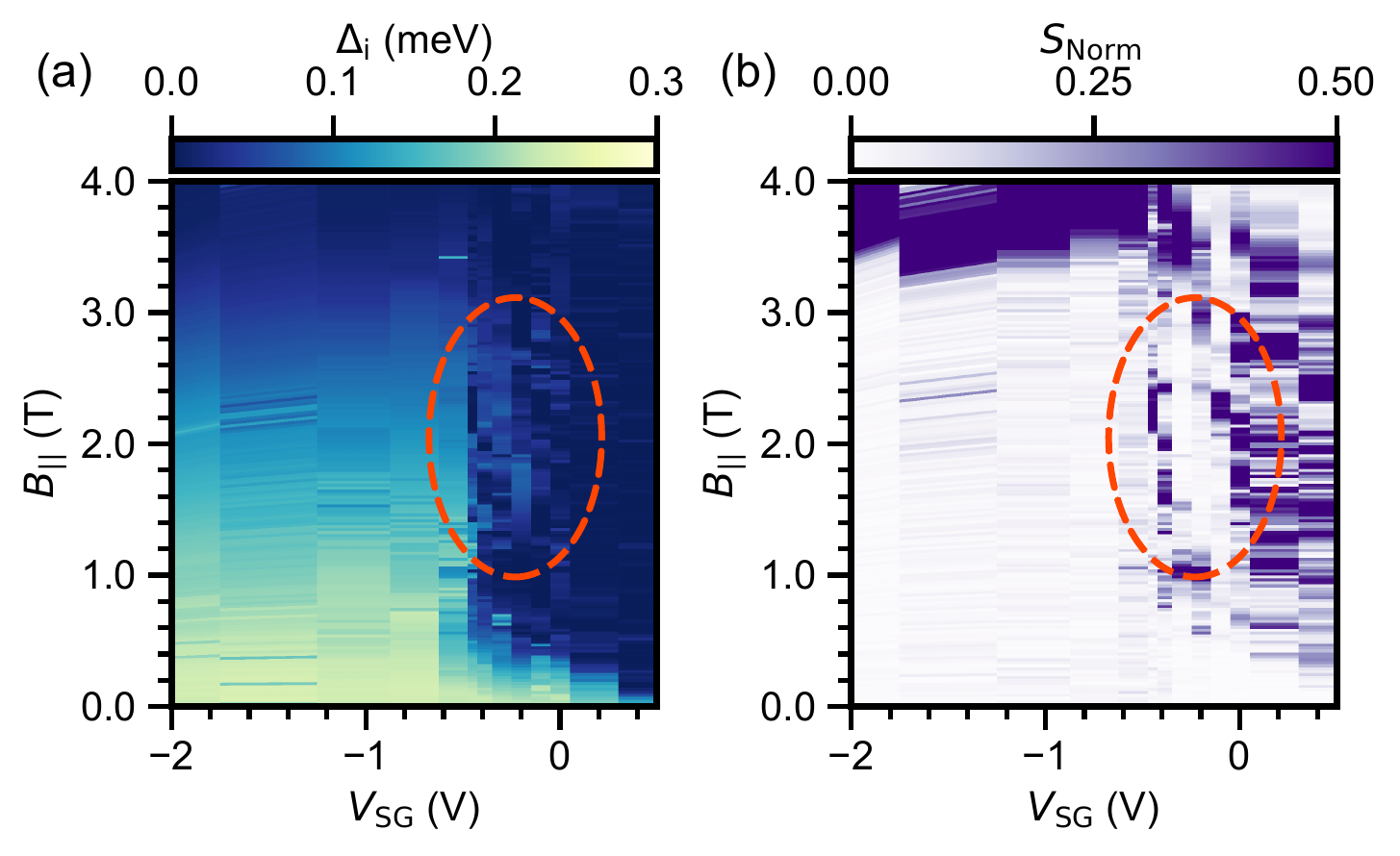}
\caption{\label{fig:4}Induced gap and nonlocal slope diagrams for device C (1\,$\mu$m long hybrid). (a) Induced gap as a function of V$_{\rm {SG}}$ and B$_{\parallel}$. (b) Normalized nonlocal slope $S_{\rm Norm}$ as a function of $V_{\rm {SG}}$ and $B_{\parallel}$. Dashed orange ellipses highlight the reopening of the induced gap.}
\end{figure}

Finally, to enhance the picture we map out the induced gap of a nanowire as a function of parallel magnetic field and super gate voltage. In Fig.~\ref{fig:4}a, we present such an induced gap diagram for the same 1\,$\mu$m long hybrid (device C). To complement this diagram, we show the corresponding normalized nonlocal slope $S_{\rm {Norm}}$ at zero bias in Fig.~\ref{fig:4}b. This quantity captures the collective behavior of the nonlocal slope from the two nonlocal signals, remaining close to zero whenever an induced gap is present in the hybrid. It is defined as $S_{\rm {Norm}} = |S_{\rm {RL}} S_{\rm {LR}}|/\sqrt{|S_{\rm {RL}} S_{\rm {LR}}|}$ where the normalization is done independently for every gate voltage. In the strong-coupling regime below $V_{\rm {SG}} < -0.5\,V$, we see that $\Delta_{\rm i}$ decays slowly when the magnetic field is increased. It closes around $B_{\rm {\parallel}}^{\rm {c}} = 3.5$\,T, which is also reflected in $S_{\rm {Norm}}$ as it deviates from zero. In contrast, above $V_{\rm {SG}} > -0.1\,V$ the semiconductor-superconductor coupling is strongly diminished which results in a significant reduction of $\Delta_{\rm i}$ and $B_{\rm {\parallel}}^{\rm {c}}$. Near the crossover between $-0.4\,V < V_{\rm {SG}} < -0.1\,V$ as indicated by the dashed orange ellipses, the closing of $\Delta_{\rm i}$ is followed by its reopening at higher magnetic fields. This is also visible in the behavior of $S_{\rm {Norm}}$, which becomes finite when the gap closes and returns to zero at the reopening. Importantly, the reopening occurs in a finite but narrow range of gate voltages. While a strong reduction of $B_{\rm {\parallel}}^{\rm {c}}$ is generically observed in our hybrids, only one out of the eleven nanowires studied in detail showed a subsequent reopening of the induced gap. In Supplementary section IV, we show phase diagrams and representative overviews of additional nanowires studied in this work.

In summary, we have demonstrated that electric fields can be used to control the bulk properties of InSb/Al/Pt three-terminal nanowires. A strong-coupling regime can be achieved, where the induced gap is large and closes only at high magnetic fields. This corresponds to a metallized nanowire which has a strong renormalization of the semiconducting properties. In contrast, a weak-coupling regime can be realized where the induced gap and critical field are strongly reduced. In long nanowires, the induced gap can be fully closed at zero magnetic field. By mapping out the induced gap diagram of a 1\,$\mu$m nanowire, we observe a closing and reopening of the induced gap in a finite range of magnetic fields and gate voltages. However, the corresponding local signals reveal an absence of zero-bias peaks. As a consequence, the gap reopening cannot be conclusively attributed to the existence of a topological phase. We speculate that the density in the hybrids is too high whenever the coupling is weakened~\cite{Antipov:2018_PRX}. In fact, it is currently unclear what are the optimal density and coupling for reaching a topological phase in InSb/Al based hybrids. Thus, a desireable future improvement would be to decouple the semiconductor and superconductor via an epitaxial barrier, such that density in the wire and the coupling could be tuned independently~\cite{Badawy_2022:AS}.

\section*{Data Availability}
Raw data presented in this work and the data processing/plotting codes are available at \url{https://doi.org/10.5281/zenodo.6913897}.

\acknowledgements{}
We thank Sebastian Heedt,Andrey Antipov, Dmitry Pikulin, Bernard van Heck, Michael Wimmer, Anton Akhmerov, Leo Bourdet and Georg W. Winkler for helpful discussions. We also thank Jan Cornelis Wolff, Mark Ammerlaan, Olaf Benningshof and Jason Mensingh for valuable technical support. This work has been financially supported by the Dutch Organization for Scientific Research (NWO), the Foundation for Fundamental Research on Matter (FOM) and Microsoft Corporation Station Q.

\section*{Author Contribution}
L.P.K, G.P.M. and N.v.L. conceived the experiment. G.P.M. and N.v.L. fabricated and measured the devices. J.Y.W., R.C.D. T.D, G.W, A.B., D.v.D, M.L and C.S. assisted with sample fabrication and/or measurements. G.P.M. and N.v.L. analyzed the transport data. G.B., S.G. and E.P.A.M.B. carried out the nanowire synthesis. G.P.M. and N.v.L. wrote the manuscript with valuable input from all authors. L.P.K. supervised the project.

\bibliography{apssamp}
\end{document}